\documentclass[journal]{IEEEtran}
\usepackage{graphicx}
\usepackage{amssymb}
\usepackage[cmex10]{amsmath}
\usepackage{bm}
\usepackage{array}
\usepackage{multirow}
\usepackage{cite}
\usepackage{url}
\usepackage{booktabs,tabularx}
\usepackage{balance}
\usepackage{stfloats}
\usepackage[caption=false,font=footnotesize]{subfig}

\usepackage[top=1.5cm,bottom=1.5cm,left=1.7cm,right=1.7cm]{geometry}

\newcommand{\re}{\operatorname{Re}}
\newcommand{\im}{\operatorname{Im}}
\newcommand{\diag}{\operatorname{diag}}

\newcommand{\Ynet}{\bm Y_{\mathrm{net}}}
\newcommand{\Yeff}{\bm Y_{\mathrm{eff}}}

\begin{document}
\bstctlcite{IEEEcontrol}

\title{DC-TORQ: Fast Assessment of Data Center-Induced Generator Torsional Oscillation Risk Quantification}

\author{{Yingyi~Tang,~\IEEEmembership{Student Member,~IEEE}, Bin~Wang,~\IEEEmembership{Senior Member,~IEEE}, Junbo~Zhao,~\IEEEmembership{Senior Member,~IEEE}, \\ Qiang Zhang,~\IEEEmembership{Senior Member,~IEEE}, Pradip Vijayan,~\IEEEmembership{Member,~IEEE}, Xiaochuan~Luo,~\IEEEmembership{Fellow,~IEEE}, \\ Tongxin Zheng,~\IEEEmembership{Fellow,~IEEE}
}
  \thanks{This work is supported by the ISO New England. Y. Tang and J. Zhao are with Thayer School of Engineering, Dartmouth College, Hanover, NH 03755 USA. B. Wang, Q. Zhang, P. Vijayan, X. Luo and T. Zheng are with ISO New England, Holyoke, MA 01040 USA.}}

\markboth{IEEE TRANSACTIONS ON POWER SYSTEMS,~2026 (Submitted)}%
{Shell \MakeLowercase{\textit{et al.}}: A Sample Article Using IEEEtran.cls for IEEE Journals}

\maketitle

\begin{abstract}
The periodic power demand of large data centers can introduce forced oscillations into the power system. When these oscillations contain frequency components near a turbine-generator torsional mode, they may excite sustained shaft torque oscillations with amplitudes exceeding the allowable cyclic torque limits. Characterizing the impact of data center power oscillations on generator shaft torque requires accounting for frequency-dependent network coupling among the data center, the transmission network, and the generators. Although electromagnetic transient (EMT) simulations can capture these interactions in detail, repeated simulations across excitation frequencies and load locations can be computationally demanding. This paper proposes an analytical framework for data center induced torsional oscillations that retains the coupled electrical and shaft dynamics. A detailed generator–network model is established, and closed-form transfer functions that map data center power perturbations to the torque in each shaft section are derived. Admissible shaft torque amplitudes are then translated into frequency-dependent limits on data center power oscillations. The proposed framework enables fast assessment of the shaft torque response while accounting for network dynamics and provides a quantitative screening criterion for candidate data center connection buses. The framework is validated on the IEEE benchmark model and applied to the IEEE 39-bus system.
\end{abstract}


\IEEEpeerreviewmaketitle

\section{Introduction}

\IEEEPARstart{L}{arge} data centers are becoming a significant source of rapidly varying demand in transmission networks. In facilities supporting artificial intelligence (AI), synchronized computing tasks can produce repeated power fluctuations that remain visible after aggregation across servers and racks \cite{choukse2025,li2024}. The North American Electric Reliability Corporation (NERC) has explicitly identified periodic workload fluctuations as a forced oscillation risk for the bulk power system \cite{nerc2025}. For turbine-generators, sustained forcing near a torsional resonance can produce large shaft torque oscillations and increase fatigue exposure.

Several organizations have developed electromagnetic transient (EMT) models to study data center integration, such as the Pacific Northwest National Laboratory (PNNL) data center model library \cite{ross2026dmlpscad}, the National Laboratory of the Rockies (NLR) detailed data center model \cite{shah2026emtdatacenters}, and the AI data center model developed by Texas A\&M University and ERCOT \cite{wang2026ups}. These models represent power conversion, controls, and load behavior at different levels of detail. Other studies address power-supply impedance and interactions within data center distribution systems \cite{sun2022part1,sun2022part2}, as well as aggregate models for transient stability studies \cite{jimenez2025,lu2026}. Alongside these modeling developments, an important concern is the potential for fluctuations in data center power demand to drive forced oscillations that affect turbine-generators in the system. If these fluctuations contain frequency components near a turbine-generator torsional mode and couple sufficiently through the electrical network, they might excite shaft torsional oscillations and amplify alternating mechanical stresses. Under unfavorable operating conditions, sustained oscillations could contribute to fatigue accumulation and shaft damages. 

Turbine-generator torsional responses depend on both shaft characteristics and interactions with the electrical network. Studies of hydropower units show that generator-to-turbine inertia ratios and shaft dynamics influence torsional stability \cite{bladh2013}. Field experience in grids containing inverter-based resources (IBRs) demonstrates that subsynchronous oscillations can propagate through the network and excite torsional modes of thermal generators, leading to protective tripping \cite{cheng2023}. More recently, subsynchronous oscillations associated with large electronic loads have been reported in ERCOT and Electranix\cite{zhang2026}, and a measured 23-Hz event has been reproduced through frequency-domain analysis and EMT simulation \cite{fan2026}.

To address concerns associated with large load power fluctuations, system operators and utilities have begun establishing limits on their magnitude and frequency content. ERCOT has proposed a 10 MW peak-to-peak limit on active power variations in the 0.1-55 Hz band, evaluated over any rolling 5~s interval \cite{ercot2026limit}. PSEG Long Island has issued requirements limiting fluctuations in the 5-55 Hz band to 3.5 MW, measured as the sum of the averaged zero-to-peak amplitudes of any two adjacent spectral bins using a rolling 10 s window \cite{pseg2026}. These limits employ distinct amplitude metrics and evaluation windows to characterize load fluctuations.

Complementary analytical and simulation-based studies assess how load fluctuations propagate through the network and affect turbine-generator shafts. Time domain simulations and transfer function analysis have been used to investigate forced oscillations, including the effects of load location, system damping, and demand harmonics \cite{valverde2025}. Allowable generator power variations have been derived from multi-mass shaft models and fatigue criteria, with load-flow-based interaction factors (IFs) used to determine admissible data center fluctuations and screen candidate locations \cite{hossain2026}. A planning-level framework combines EMT case studies of shaft torque amplification with endurance-based screening criteria for non-resonant conditions and short-circuit-based estimates of generator load sharing, while emphasizing the need for unit-specific assessment near torsional modal frequencies \cite{epri2026}. For hydropower units, EMT frequency scans have been used to evaluate network propagation and plant torque amplification and relate the resulting shaft stresses to fatigue margins. These results highlight the influence of the generator-to-turbine inertia ratio and torsional damping \cite{chatterjee2026}.

These studies provide practical approaches to assessing torsional risk. While the load-flow-based interaction factors in \cite{hossain2026} and the short-circuit-based load sharing estimates in \cite{epri2026} provide efficient screening measures, assessing frequency-dependent load-to-generator responses requires additional consideration of the coupled dynamics of the electrical network and turbine-generator shafts. EMT frequency scans, as used in \cite{chatterjee2026}, can capture these effects, but generally require repeated simulations across excitation frequencies and system conditions change. The computational burden could become substantial when a large number of load locations and network configurations must be evaluated.

Motivated by these challenges, this paper proposes an analytical framework for data center induced torsional oscillations of turbine-generators that retains the coupled electrical and shaft dynamics and yields closed-form transfer functions from the data center power fluctuation to the shaft torques. This enables rapid evaluation of whether the shaft torques exceed admissible limits for arbitrary data center connection buses without resorting to EMT simulation. The key contributions of this paper are summarized as follows.

1) A coupled generator–network model of power systems with data center loads is established, capturing turbine-generator multi-mass shaft dynamics and network dynamics. Black-box IBRs are also considered through their terminal admittances, which extends the generality of the proposed model.

2) Transfer functions from a data center power perturbation at a candidate bus to the generator states and shaft torques are derived for the coupled system. Harmonic superposition then yields the shaft response to arbitrary periodic load profiles, which allows us to analytically assess the oscillations.

3) Admissible shaft torque amplitudes are translated into frequency-dependent limits on data center power oscillations, providing a quantitative screening criterion for sustained oscillations.

\section{Generator and Shaft Dynamics}
This section provides some basic information on generators, the multi-mass shaft model, as well as the small-signal analysis model.

\subsection{Electrical Model With Stator-Flux Dynamics}
\begin{figure}[!t]
\centering
\includegraphics[width=3.4in]{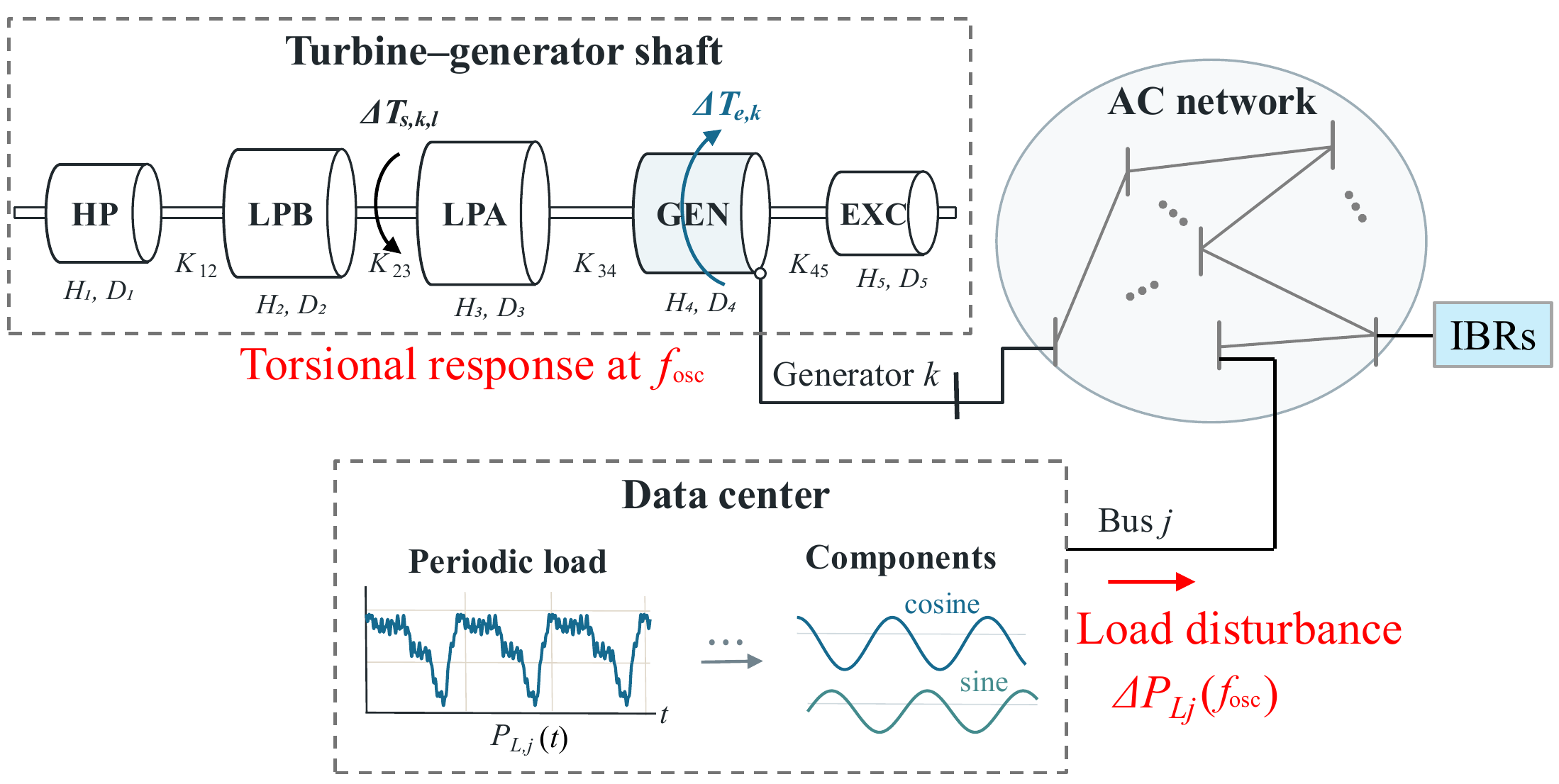}
\vspace{-0.3cm}
\caption{Schematic of the turbine-generator torsional response to a periodic data center load disturbance.}
\vspace{-0.3cm}
\label{fig:Datacenter}
\end{figure}

The studied system is shown in Fig. \ref{fig:Datacenter} where a data center load oscillation at Bus $j$ propagates through the AC network to generator $k$. Each of the $N$ turbine-generators is represented by the GENDCO model \cite{psse,ieee1110}. The form used here assumes equal direct- and quadrature-axis subtransient reactances, $X''_d=X''_q=X''$. With the generator index omitted for brevity, the electrical equations in the rotor $dq$ frame are given by
\begin{equation}
\begin{cases}
\dot E'_{q}=(E_{fd,0}-X_{ad}I_{fd})/T'_{d0},\\
\dot E'_{d}=-X_{aq}I_{lq}/T'_{q0},\\
\dot\psi_{kd}=[E'_q-\psi_{kd}-(X'_d-X_l)i_d]/T''_{d0},\\
\dot\psi_{kq}=[E'_d-\psi_{kq}+(X'_q-X_l)i_q]/T''_{q0},\\
\dot\psi_{d}=\omega_{b}\bigl[u_{d}+(1+\omega_{g})\,\psi_{q}+R_{a}i_{d}\bigr],\\
\dot\psi_{q}=\omega_{b}\bigl[u_{q}-(1+\omega_{g})\,\psi_{d}+R_{a}i_{q}\bigr].
\end{cases}
\label{eq:gendco}
\end{equation}

The stator currents and electromagnetic torque satisfy
\begin{equation}
i_d=(\psi''_d-\psi_d)/X'',\quad
i_q=(-\psi''_q-\psi_q)/X'',
\label{eq:gendco-alg}
\end{equation}
\begin{equation}
    T_e =\psi_di_q-\psi_qi_d,
\label{eq:gendco-alg2}
\end{equation}
where
\begin{equation}
\begin{cases}
\psi''_d=\frac{(X''-X_l)E'_q+(X'_d-X'')\psi_{kd}}{X'_d-X_l},\\
\psi''_q=\frac{(X''-X_l)E'_d+(X'_q-X'')\psi_{kq}}{X'_q-X_l}.
\end{cases}
\end{equation}
Here $E'_q$, $E'_d$ are the transient voltages, $\psi_{kd}$,
$\psi_{kq}$ are the damper-winding flux linkages, $\psi_d$, $\psi_q$ are the stator flux linkages, and $\psi''_d$, $\psi''_q$ are the corresponding subtransient flux linkages. The terminal voltage and current components are $u_d,u_q$ and $i_d,i_q$. $I_{fd}$ is the field current and $I_{lq}$ is the current in the $q$-axis transient rotor circuit. The field voltage $E_{fd,0}$ is fixed at its operating point value, and $\omega_g$ is the generator-mass speed
deviation in per unit. The base angular frequency is $\omega_b=2\pi f_0$.

The reactances $X_d,X_q$, $X'_d,X'_q$, and $X_l$ are the synchronous, transient, and leakage reactances, respectively, with $X_{ad}=X_d-X_l$ and $X_{aq}=X_q-X_l$. The armature resistance is $R_a$. $T'_{d0},T'_{q0}$ and $T''_{d0},T''_{q0}$ are the open-circuit transient and subtransient time constants. 

In \eqref{eq:gendco}, unlike GENROU and other standard stability models, the stator flux linkage dynamics ($\dot{\psi}_d$, $\dot{\psi}_q$) are retained to include the impact of the DC offset torque components \cite{segerstrom2023}. Accordingly, the electrical states for the $k$-th generator can be stacked as
$\bm x_{e,k}=[E'_q\ \ E'_d\ \ \psi_{kd}\ \ \psi_{kq}\ \ \psi_d\ \
\psi_q]^{\top}$.

\subsection{Multi-Mass Shaft Model}
Consider generator $k$ with $n_k$ rotating masses connected by $n_k-1$ shaft sections. Let $\bm\theta_k,\bm\omega_k\in\mathbb{R}^{n_k}$ collect the rotor angles and per-unit speed deviations of the masses. Let $\bm e_k\in\mathbb R^{n_k}$ denote the unit vector with a one at the generator-mass position and zeros elsewhere. The multi-mass shaft model dynamics can be expressed as
\begin{equation}
\begin{cases}
\dot{\bm\theta}_k=\omega_b\,\bm\omega_k,\\
\bm M_k\dot{\bm\omega}_k=\bm T_{m,k}-\bm e_k T_{e,k}
-\bm B_k^{\top}\bm T_{s,k}-\bm D_k\bm\omega_k ,
\end{cases}
\label{eq:shaft-odes}
\end{equation}
where $\bm T_{m,k}$ is the vector of applied mechanical torques and $\bm T_{s,k}$ contains the shaft-section torques. Define the incidence matrix
\begin{equation}
\bm B_k=
\begin{bmatrix}
1&-1&&\\
&\ddots&\ddots&\\
&&1&-1
\end{bmatrix}
\in\mathbb R^{(n_k-1)\times n_k}.
\label{eq:shaft_incidence}
\end{equation}
Then
\begin{equation}
\begin{cases}
\bm T_{s,k}=\bm K_{s,k}\bm B_k\bm\theta_k,\\
\bm K_{s,k}=\diag(K_{k,1},\ldots,K_{k,n_k-1}),\\
\bm M_k=2\diag(H_{k,1},\ldots,H_{k,n_k}),\\
\bm D_k=\diag(D_{k,1},\ldots,D_{k,n_k}).
\end{cases}
\label{eq:shaft-def}
\end{equation}
The stiffness of shaft section $\ell$ is $K_{k,\ell}$, while $H_{k,i}$ and $D_{k,i}$ are the inertia constant and self-damping coefficient of mass $i$. The sign convention gives $T_{s,k,\ell}=K_{k,\ell}(\theta_{k,i}-\theta_{k,i+1})$.

\subsection{Small-Signal Model}
By stacking the electrical and mechanical states of all generators as
\begin{equation}
\bm x=\bigl[\bm x_{e,1}^{\top}\ \ \bm\theta_1^{\top}\ \ \bm\omega_1^{\top}
\ \ \cdots\ \ \bm x_{e,N}^{\top}\ \ \bm\theta_N^{\top}\ \
\bm\omega_N^{\top}\bigr]^{\top},
\label{eq:state}
\end{equation}
the multi-mass generators' dynamics and terminal current outputs in \eqref{eq:gendco}--\eqref{eq:shaft-def} can be expressed as
\begin{equation}
\dot{\bm x}=\bm f(\bm x,\bm v_G), \quad ,\bm i_G=\bm h_I(\bm x) 
\label{eq:nonlin}
\end{equation}
where $\bm v_G,\bm i_G\in\mathbb{R}^{2N}$ are the generator terminal voltages and injected currents in the common network $xy$ frame. For each generator,
\begin{equation}
    \bm v_{G,k}=\bm P(\theta_{k,gen})\begin{bmatrix}u_{d,k} \\ u_{q,k}\end{bmatrix},  \quad  \bm i_{G,k}=\bm P(\theta_{k,gen})\begin{bmatrix}i_{d,k} \\ i_{q,k}\end{bmatrix},
\end{equation}
with 
\begin{equation}
    \bm P(\theta)\triangleq 
\begin{bmatrix}\sin\theta&\cos\theta\\ 
-\cos\theta&\sin\theta\end{bmatrix}
\end{equation}
The angle $\theta_{k,gen}$ locates the generator mass's $q$ axis in the network frame. 

Linearizing \eqref{eq:nonlin} around the operating point
$(\bm{x}_0, \bm{v}_{G,0})$ yields the small-signal model of the generators
\begin{equation}
\begin{aligned}
&\Delta\dot{\bm x}=\bm A_0\Delta\bm x+\bm B_u\Delta\bm v_G,\\
&\Delta\bm i_G=\bm C_I\Delta\bm x,
\end{aligned}
\end{equation}
where
\begin{equation}
\bm A_0=\left.\frac{\partial\bm f}{\partial\bm x}\right|_{0},\quad
\bm B_u=\left.\frac{\partial\bm f}{\partial\bm v_G}\right|_{0},\quad
\bm C_I=\left.\frac{\partial\bm h_I}{\partial\bm x}\right|_{0}.
\label{eq:jac}
\end{equation}

Assuming zero initial perturbations, the Laplace-domain representation is given by
\begin{equation}
\begin{aligned}
&\Delta\bm x(s)=(s\bm I-\bm A_0)^{-1}\bm B_u\Delta\bm v_G(s),\\
&\Delta\bm i_G(s)=\bm C_I\Delta\bm x(s).
\end{aligned}
\label{eq:iGVG}
\end{equation}

\section{Derivation of the Generator Response Model to Data Center Demand}
\label{sec:net}
\subsection{Periodic Data Center Demand Model}
Consider a data center connected to Bus $j$. Its active power demand is expressed as the sum of a steady-state component and a small periodic perturbation as
\begin{equation}
\begin{aligned}
        & P_{L,j}(t)=P_{L,j0}+\Delta P_{L,j}(t),\\
    & \Delta P_{L,j}(t)=\sum_{h=1}^{\infty}a_h\cos \big(h\omega_\mathrm{osc} t+\psi_h\big).
\end{aligned}
\end{equation}
Here $P_{L,j0}$ is the steady-state active power demand.
$a_h$ and $\psi_h$ are the peak amplitude and phase angle of the $h$-th harmonic, respectively. The fundamental angular frequency is $\omega_f=2\pi f_\mathrm{osc}$. This representation covers sinusoidal forcing and periodic workload profiles with any specified harmonics \cite{valverde2025}.

The data center is assumed to operate at a constant power factor:
\begin{equation}
S_{L,j}(t)=(1+j\kappa)P_{L,j}(t),\qquad Q_{L,j}(t)=\kappa P_{L,j}(t),
\label{eq:load_power_factor}
\end{equation}
where $\kappa$ is the reactive-to-active-power ratio. With $\kappa=0$, the data center operates at unity power factor with zero reactive power demand.

Taking current injection into the grid as positive, the current injected by the data center is given by
\begin{equation}
I_{L,j}(t)=-\frac{S^{*}_{L,j}(t)}{V^{*}_{L,j}(t)}=-\frac{(1-j\kappa)P_{L,j}(t)}{V^{*}_{L,j}(t)},
\label{eq:Idc}
\end{equation}
Linearizing \eqref{eq:Idc} about the operating point yields
\begin{equation}
\Delta I_{L,j}(t)=b_{L,j}\,\Delta P_{L,j}(t)+a_{L,j}\,\Delta V^{*}_{L,j}(t),
\label{eq:dataI}
\end{equation}
where 
\begin{equation}
    b_{L,j}=-\frac{(1-j\kappa)}{V^{*}_{L,j,0}}, \quad a_{L,j}=\frac{S^{*}_{L,j,0}}{\bigl(V^{*}_{L,j,0}\bigr)^{2}}.
\end{equation}

Define rectangular-coordinate perturbations as
\begin{equation}
    \Delta \bm v_{L,j}(t)=\begin{bmatrix}
        \Delta V^{x}_{L,j}(t)\\\Delta V^{y}_{L,j}(t)
    \end{bmatrix}, \quad \Delta \bm i_{L,j}(t)=\begin{bmatrix}
        \Delta I^{x}_{L,j}(t)\\\Delta I^{y}_{L,j}(t) \end{bmatrix},
\end{equation}
Then the data center current in \eqref{eq:dataI} can be represented in the Laplace domain as
\begin{equation}
\Delta\bm i_{L,j}(s)
=\bm J_{L,j}\,\Delta\bm v_{L,j}(s)
+\mathcal{R}(b_{L,j})\,\Delta P_{L,j}(s)   
\label{eq:idcvec}
\end{equation}
where
\begin{equation}
\bm J_{L,j}=
\begin{bmatrix}
\re a_{L,j} & \im a_{L,j}\\
\im a_{L,j} & -\re a_{L,j}
\end{bmatrix},\ 
\mathcal{R}(b_{L,j})=
\begin{bmatrix}\re b_{L,j}\\ \im b_{L,j}\end{bmatrix}.
\label{eq:Jdc}
\end{equation}
It can be seen from \eqref{eq:idcvec} that the change in the data center terminal current is determined by both the variation in its active power demand and the corresponding variation in terminal voltage.

\subsection{Consideration of Network Dynamics}
The torsional modes of interest lie in the subsynchronous range, where network inductances and capacitances affect how the data center disturbance propagates to the generator terminals. A quasi-steady-state phasor model neglects these electromagnetic dynamics and therefore does not fully capture this frequency-dependent coupling. Accordingly, the network is instead modeled dynamically in a common synchronously rotating $xy$ frame. Let $\Ynet(s)$ be the $2M\times2M$ nodal admittance matrix of the $M$-bus network in the common $xy$ frame, and let $\Delta\bm v=[\Delta\bm v_1,\ldots,\Delta\bm v_M)]^\top$. Unlike the constant impedance and admittance blocks used in a quasi-steady-state model, the branch and shunt blocks in this formulation depend on $s$. For example, a series RL branch and a parallel RC shunt have the blocks
\begin{equation}
\begin{aligned}
\bm Z_{RL}(s)&=
\begin{bmatrix}R+sL&-\omega_bL\\\omega_bL&R+sL\end{bmatrix},\\
\bm Y_{RC}(s)&=
\begin{bmatrix}1/R+sC&-\omega_bC\\\omega_bC&1/R+sC\end{bmatrix}.
\end{aligned}
\label{eq:network_blocks}
\end{equation}
These blocks describe perturbations about the synchronous operating point and $s$ is the perturbation-frequency variable. Ordinary loads represented by fixed RLC elements are included in $\Ynet$.

Define the selection matrices $\bm E_{L,j}\in\mathbb{R}^{2M\times2}$ and $\bm E_G\in\mathbb{R}^{2M\times2N}$, whose columns are the unit vectors of the coordinates of data center Bus $j$ and of the $N$ generator buses, respectively. Consequently, $\bm E_{L,j}\Delta\bm i_{L,j}(s)$ and $\bm E_G\Delta\bm i_G(s)$ places the data center and generator currents at their respective buses. Then the network equation in $s$-domain is
\begin{equation}
\Ynet(s)\Delta\bm v(s)=\bm E_G\Delta\bm i_G(s)+\bm E_{L,j}\Delta\bm i_{L,j}(s).
\label{eq:nodal}
\end{equation}

Applying $\Delta\bm v_{L,j}(s)=\bm E_{L,j}^{\top}\Delta\bm v(s)$ and \eqref{eq:idcvec}, define 
\begin{equation}
\Yeff(s)\triangleq\Ynet(s)-\bm E_{L,j}\bm J_{L,j}\bm E_{L,j}^{\top}.
\label{eq:Yeff}
\end{equation}
The network equation \eqref{eq:nodal} becomes
\begin{equation}
\Yeff(s)\Delta\bm v(s)
=\bm E_G\Delta\bm i_G(s)
+\bm E_{L,j}\mathcal R(b_{L,j})\Delta P_{L,j}(s).
\label{eq:nodal2}
\end{equation}

Substituting $\Delta\bm v_G(s)=\bm E_G^{\top}\Delta\bm v(s)$ into \eqref{eq:nodal2} yields 
\begin{equation}
  \Delta\bm v_G(s)=\bm W(s) \Delta\bm i_G(s)+\bm z_{L,j}(s)\,\Delta P_{L,j}(s),
\label{eq:vGnet}  
\end{equation}
where
\begin{equation}
  \begin{cases}
      \bm W(s)=\bm E_G^{\top}\Yeff^{-1}(s)\,\bm E_G,\nonumber\\
      \bm z_{L,j}(s)=\bm E_G^{\top}\Yeff^{-1}(s)\,
\bm E_{L,j}\mathcal{R}(b_{L,j}).
  \end{cases}  
\end{equation}
Here $\bm W(s)$ is the network impedance matrix at the generator terminals, and $\bm z_{L,j}(s)$ maps the load-power perturbation to terminal voltage with generator current perturbations held fixed. 

Combining \eqref{eq:vGnet} and \eqref{eq:iGVG} yields,
\begin{equation}
\bigl[s\bm I-\bm A_0-\bm B_u\bm W(s)\bm C_I\bigr]\Delta\bm x(s)
=\bm B_u\bm z_{L,j}(s)\,\Delta P_{L,j}(s).
\label{eq:closed}
\end{equation}
Accordingly, the transfer function from the data center active power oscillation to all generator states can be defined as
\begin{equation}
\begin{aligned}
\bm H_{xP_{L,j}}(s)&\triangleq\frac{\Delta\bm x(s)}{\Delta P_{L,j}(s)}\\
&=\bigl[s\bm I-\bm A_0-\bm B_u\bm W(s)\bm C_I\bigr]^{-1}\bm B_u\,
\bm z_{L,j}(s)
\end{aligned}
\label{eq:HxP}
\end{equation}


\subsection{Extension to Black-Box IBR Models}
An IBR whose internal controller model is represented by black-box simulation model can still be included in the framework through a terminal admittance model identified from its frequency response, for example by frequency scanning \cite{dsatools} or impedance measurement \cite{sun2011,cespedes2014} followed by vector fitting \cite{gustavsen1999}. Let $\mathcal B$ denote the set of such units. With current injection into the network taken as positive, the terminal relation of unit $r\in\mathcal B$ is
\begin{equation}
\Delta\bm i_r(s)
={\bm Y}_r(s)\Delta\bm v_r(s),
\qquad
\Delta\bm v_r(s)=\bm E_r^{\top}\Delta\bm v(s),
\label{eq:ibr_terminal}
\end{equation}
where $\bm E_r\in\mathbb R^{2M\times2}$ selects the connection bus of the unit and ${\bm Y}_r(s)$ is the fitted admittance expressed in the common $xy$ frame.

Substituting \eqref{eq:ibr_terminal} into the nodal equation yields
\begin{equation}
\widehat{\bm Y}_{\mathrm{eff}}(s)\Delta\bm v(s)=\bm E_G\Delta\bm i_G(s)+\bm E_{L,j}\mathcal R(b_{L,j})\Delta P_{L,j}(s),
\label{eq:ibr_effective_network}
\end{equation}
where
\begin{equation}
\widehat{\bm Y}_{\mathrm{eff}}(s)=\Yeff(s)
-\sum_{r\in\mathcal B}
\bm E_r{\bm Y}_r(s)\bm E_r^{\top}.
\label{eq:ibrYeff}
\end{equation}
Replacing $\Yeff$ by $\widehat{\bm Y}_{\mathrm{eff}}$ in \eqref{eq:ibrYeff} then gives the response of the generators that remain explicitly modeled. If a fitted terminal model replaces a unit that was previously represented explicitly, the states and current-injection block of that unit would be removed from $\bm A_0$, $\bm B_u$, $\bm C_I$, and $\bm E_G$ so that its response is not counted twice.

The extension thus allows IBRs to be included as long as their simulation models are available such as black-box models, while still capturing the influence of their terminal dynamics on data center power oscillation propagation and on the torque response of the turbine-generators.

\section{Shaft Torque and Load Oscillation Limit}
\subsection{Torsional Modes and Shaft Torque Response}
\begin{figure}[!t]
\centering
\includegraphics[width=3.4in]{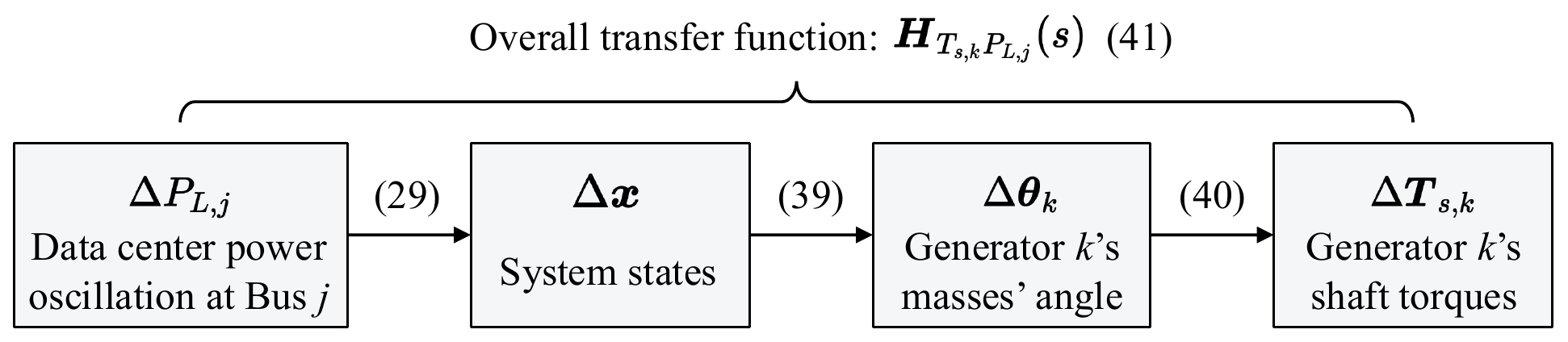}
\vspace{-0.3cm}
\caption{Transfer-function mapping from data center power oscillations at Bus $j$ to shaft torque deviations of generator $k$.}
\vspace{-0.3cm}
\label{fig:map}
\end{figure}
For the standalone generator-turbine shaft system (i.e., neglecting electrical network interactions), linearizing \eqref{eq:shaft-odes} with $\Delta\bm T_{m,k}=\bm 0$ and
eliminating $\Delta\bm\omega_k=(s/\omega_b)\Delta\bm\theta_k$ gives
\begin{equation}
  \bm Z_{m,k}(s)\,\Delta\bm\theta_k(s)=-\bm e_k\,\Delta T_{e,k}(s) , 
\end{equation}
where
\begin{equation}
\bm Z_{m,k}(s)=\bm B_k^{\top}\bm K_{s,k}\bm B_k
+\frac{s}{\omega_b}\bm D_k+\frac{s^2}{\omega_b}\bm M_k.
\label{eq:Zm}
\end{equation}
The shaft torque deviations can be obtained via
\begin{equation}
    \Delta\bm T_{s,k}=\bm K_{s,k}\bm B_k\Delta\bm\theta_k
    \label{eq:Ts_alone}
\end{equation}

Combing \eqref{eq:Zm} and \eqref{eq:Ts_alone}, the transfer function characterizing shaft torque amplification in section $\ell$ of generator $k$ is
\begin{equation}
G_{T_{s,\ell}T_{e,k}}(s)\triangleq\frac{\Delta T_{s,k,\ell}(s)}{\Delta T_{e,k}(s)}
=-\bm e_\ell^{\top}\bm K_{s,k}\bm B_k\,\bm Z_{m,k}^{-1}(s)\,\bm e_k,
\label{eq:GTsTe}
\end{equation}
where $\bm e_\ell\in\mathbb{R}^{n_k-1}$ selects the shaft section. The oscillatory torsional modes of the mechanical subsystem are associated with the nonreal roots of
\begin{equation}
\det\bm Z_{m,k}(s)=0
\label{eq:modes}
\end{equation}
These modes involve relative angular motion between the shaft masses. 

In the interconnected system, electromagnetic torque feedback couples the shaft motion to the electrical dynamics and modifies the frequencies and damping of the corresponding modes. Based on \eqref{eq:closed}, setting the external power perturbation to zero gives
\begin{equation}
\begin{aligned}
\left[\lambda_m\bm I-\bm A_0-\bm B_u\bm W(\lambda_m)\bm C_I\right]\bm r_m
=\bm 0, \quad \bm r_m\neq\bm 0,\\
\Rightarrow \det[\lambda_m\bm I-\bm A_0-\bm B_u\bm W(\lambda_m)\bm C_I]=0,
\end{aligned}
\label{eq:coupled_modal}
\end{equation}
where $\bm r_m$ contains the components of the system torsional mode shape associated with $\bm x$. The network dynamics are retained through $\bm W(s)$.

To evaluate the shaft torque responses of generator $k$ to the data center power perturbation, define $\bm E_{\theta,k}$ as the matrix that selects its angle states from the full state vector in \eqref{eq:state}
\begin{equation}
\Delta\bm\theta_k=\bm E_{\theta,k}\Delta\bm x.
\end{equation}
Then based on \eqref{eq:Ts_alone}, the shaft torque responses are obtained as
\begin{equation}
    \Delta\bm T_{s,k}= \bm C_{T_{s,k}}\Delta\bm x, \quad \bm C_{T_{s,k}}\triangleq\bm K_{s,k}\bm B_k\bm E_{\theta,k}.
\label{eq:CTsx}
\end{equation}

Combining \eqref{eq:HxP} and \eqref{eq:CTsx} gives the closed-form transfer function from the Bus $j$ data center power perturbation to the generator $k$'s shaft torques:
\begin{equation}
    \bm H_{T_{s,k}P_{L,j}}(s)\triangleq\bm C_{T_{s,k}}\bm H_{xP_{L,j}}(s).
\label{eq:HTsP}
\end{equation}
where the $\ell$-th row of $\bm H_{T_{s,k}P_{L,j}}(s)$ specifies the torque in shaft section $\ell$:
\begin{equation}
H_{T_{s,k,\ell}P_{L,j}}(s)=\bm e_\ell^{\top}\bm H_{T_{s,k}P_{L,j}}=\frac{\Delta T_{s,k,\ell}(s)}{\Delta P_{L,j}(s)}.
\label{eq:H_ell}
\end{equation}

As illustrated in Fig. \ref{fig:map}, data center power oscillations at Bus $j$ propagate through the system states to the mass angle and shaft torque deviations of generator $k$. Combining these mappings yields the overall transfer function $\bm H_{T_{s,k}P_{L,j}}(s)$ in \eqref{eq:HTsP}.

For a stable operating point, the steady-state peak amplitude of the shaft torque oscillation induced by the $h$-th harmonic of the data-center demand is
\begin{equation}
|\Delta\tilde T_{s,k,\ell,h}|=|H_{T_{s,k,\ell}P_{L,j}}(j h\omega_f)|a_h.
\label{eq:Tsh}
\end{equation}
The total steady-state torque response is obtained by superposing the harmonic responses as
\begin{equation}
\Delta T_{s,k,\ell}^{\mathrm{ss}}(t)
=\operatorname{Re}
\left\{
\sum_{h=1}^{\infty}
H_{T_{s,k,\ell}P_{L,j}}({j}h\omega_f)
a_h e^{\mathrm{j}(h\omega_f t+\psi_h)}
\right\}.
\label{eq:Ts_superposition}
\end{equation}

It can be interpreted from \eqref{eq:HTsP} that, the shaft torque response depends both on the system modes in \eqref{eq:coupled_modal} and data center oscillation $\Delta P_{L,j}(s)$. A harmonic of the data center demand can strongly excite one of the torsional modes determined by \eqref{eq:coupled_modal} when its frequency approaches the modal frequency, provided that the mode is lightly damped. 

The resulting torque amplitude may exceed the admissible limit associated with the shaft section's material endurance criterion even when the operating point remains stable. To assess this risk at the planning stage,  \eqref{eq:HTsP}– \eqref{eq:Ts_superposition} provide a fast means of evaluating the shaft torque response to multiharmonic data center load fluctuations.

\subsection{Torque-Based Data Center Power Oscillation Limits}
Following \cite{zhang2026} and \cite{hossain2026}, let $EL_{\mathrm{set},k,\ell}$ be the admissible alternating torque amplitude of section $\ell$ of generator $k$. For a specified shaft geometry, alternating shear stress is proportional to torque. A material endurance limit can therefore be converted to a torque threshold when geometry, stress
concentrations, and mean-stress effects are known. Otherwise, the threshold could be a prescribed fatigue limit.

For an admissible shaft torque threshold $EL_{\mathrm{set},k,\ell}$, the admissible data center power oscillation amplitude at frequency $f$ can be determined based on \eqref{eq:Tsh} as
\begin{equation}
\Delta P^{\mathrm{EX}}_{L,j,\mathrm{lim},k}(f) =
\min_\ell\frac{EL_{\mathrm{set},k,\ell}}{\bigl|H_{T_{s,k,\ell}P_{L,j}}(j 2\pi f)\bigr|}.
\label{eq:limEX_section}
\end{equation}
This quantity is the admissible amplitude of the data center power oscillation at Bus $j$ with respect to generator $k$.

The system-wide limit at Bus $j$ is obtained by taking the minimum over all generators:
\begin{equation}
\Delta P^{\mathrm{EX}}_{L,j,\mathrm{lim}}(f)
=\min_k\Delta P^{\mathrm{EX}}_{L,j,\mathrm{lim},k}(f).
\label{eq:limEX_system}
\end{equation}

Consequently, for a periodic data center demand, a sufficient, conservative limit can be determined as
\begin{equation}
\sum_h |H_{T_{s,k,\ell}P_{L,j}}(j h\omega_f)|a_h \le EL_{\mathrm{set},k,\ell}\quad\text{for all }k,\ell.
\label{eq:combined_torque_bound}
\end{equation}
Notably, this bound limits the peak amplitude of the alternating shaft torque. Estimating cumulative fatigue damage requires a separate assessment of the loading history and material fatigue properties.

\section{Numerical Results and Analysis}
The proposed framework is first assessed using a single-machine benchmark system and then applied to a modified IEEE 39-bus system. A third study incorporates a black-box IBR through its terminal admittance obtained via frequency scan. 

\subsection{Benchmark Validation and Data Center Load Integration}
\subsubsection{Validation}
\begin{figure}[!t]
\centering
\vspace{-0.1cm}
\includegraphics[width=3.4in]{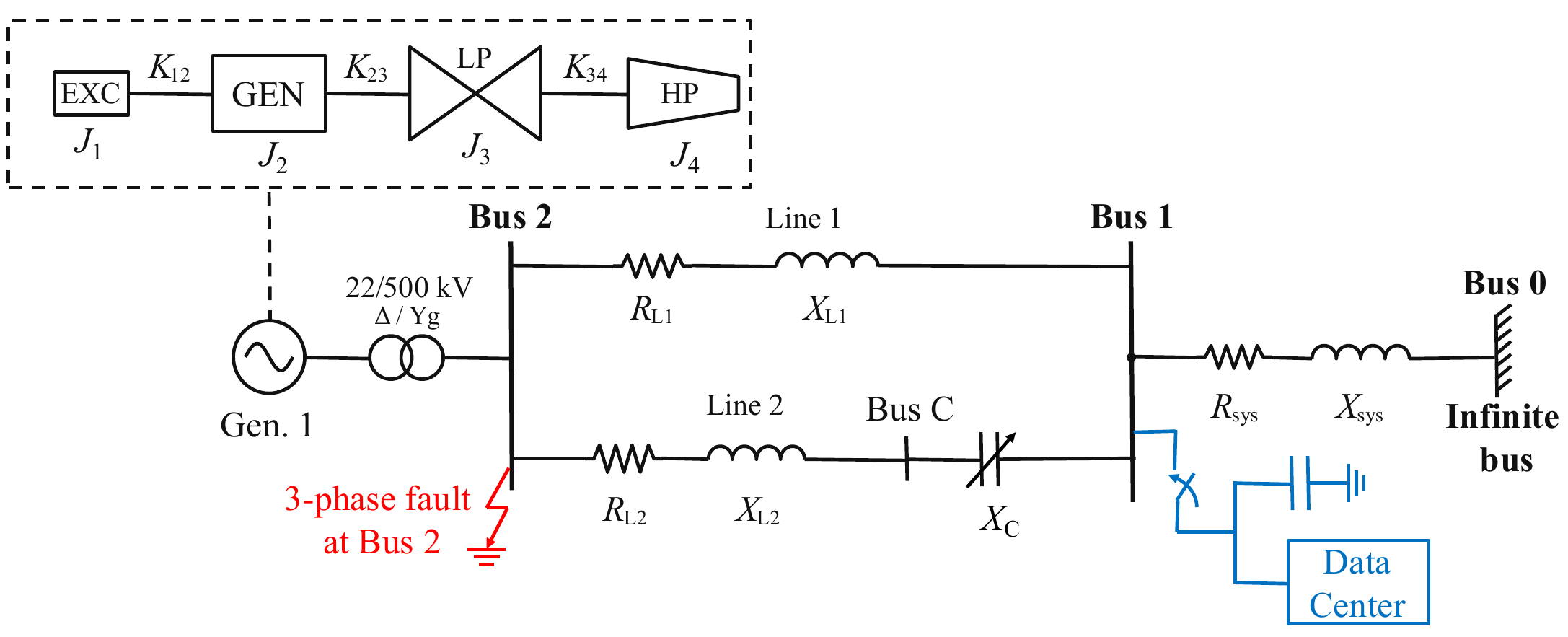}
\vspace{-0.3cm}
\caption{IEEE second benchmark model for SSR \cite{sbm1985}.}
\vspace{-0.3cm}
\label{fig:SBM_sys}
\end{figure}

As shown in Fig. \ref{fig:SBM_sys}, the proposed framework is validated on System 1 of the IEEE second benchmark model for subsynchronous resonance (SSR) analysis \cite{sbm1985}, with the generator and system parameters following the benchmark specification. 

\begin{figure}[!t]
\centering
\vspace{-0.1cm}
\includegraphics[width=3.3in]{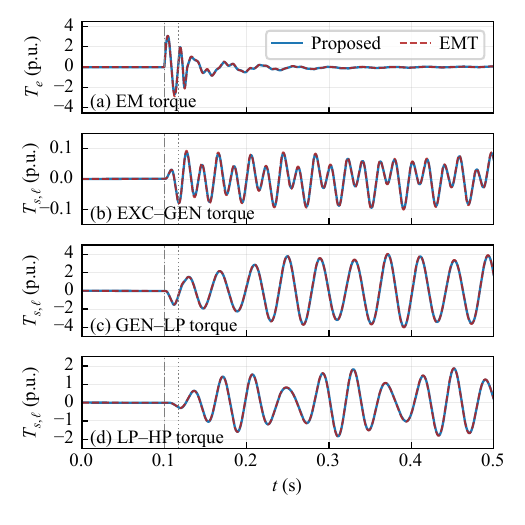}
\vspace{-0.3cm}
\caption{Torque response after 0.017 s fault clearing time for Case 1A in \cite{sbm1985}.}
\vspace{-0.3cm}
\label{fig:SBM_fault}
\end{figure}

Table \ref{tab:SBM_validation} (a) compares the derived undamped torsional frequencies and generator-referred modal inertia constants with the published values. To validate the transient torque response, Case 1A of \cite{sbm1985} is reproduced, in which a three-phase fault is applied at Bus 2 and cleared after 0.017 s. Table \ref{tab:SBM_validation} (b) reports the resulting shaft torques obtained from the proposed framework, the EMT simulation, and the benchmark, while Fig. \ref{fig:SBM_fault} overlays the electromagnetic and shaft torque waveforms from the proposed framework and the EMT simulation. The close agreement in modal quantities, torque amplitudes, and transient waveforms verifies the implementation of the generator and multi-mass shaft models.

\begin{table}[t]
\centering
\caption{Validation Results for the IEEE Second Benchmark System}
\vspace{-0.3cm}
\label{tab:SBM_validation}
\setlength{\tabcolsep}{3pt}
(a) Mechanical modal frequencies and inertias

\vspace{3pt}

\begin{tabular*}{\columnwidth}
{@{\extracolsep{\fill}}ccccc@{}}
\toprule
& \multicolumn{2}{c}{Frequency (Hz)}
& \multicolumn{2}{c}{Modal inertia (s)} \\
\cmidrule(lr){2-3}\cmidrule(lr){4-5}
Mode
& Proposed & \cite{sbm1985}
& Proposed & \cite{sbm1985} \\
\midrule
1 & 24.6367 & 24.65 & 1.5505  & 1.55 \\
2 & 32.3737 & 32.39 & 9.4057  & 9.39 \\
3 & 51.1092 & 51.10 & 73.6766 & 74.8 \\
\bottomrule
\end{tabular*}

\vspace{8pt}

(b) Shaft torque responses for \cite{sbm1985}'s Case 1A
\vspace{3pt}

\begin{tabular*}{\columnwidth}
{@{\extracolsep{\fill}}cccc@{}}
\toprule
& \multicolumn{3}{c}{Half peak-to-peak torque (p.u.)} \\
\cmidrule(lr){2-4}
Shaft section & Proposed & EMT & \cite{sbm1985} \\
\midrule
EXC--GEN & 0.097495 & 0.094303 & 0.101 \\
GEN--LP  & 4.049111 & 3.990952 & 4.020 \\
LP--HP   & 1.869017 & 1.849591 & 1.970 \\
\bottomrule
\end{tabular*}
\end{table}

\begin{figure}[!t]
\centering
\vspace{-0.1cm}
\includegraphics[width=3.3in]{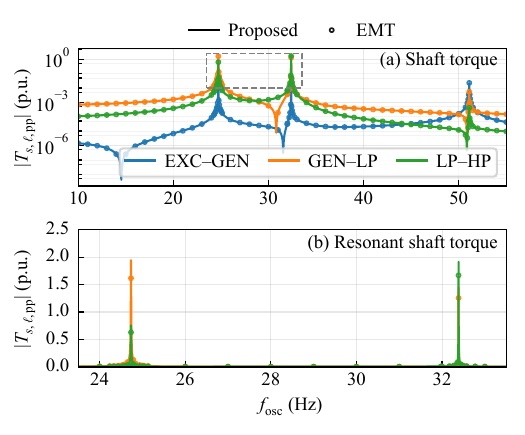}
\vspace{-0.3cm}
\caption{Generator shaft torque response with added data center forced oscillation over $f_\mathrm{osc}$.}
\vspace{-0.3cm}
\label{fig:SBM_dc}
\end{figure}

\subsubsection{Response to Data Center Power Oscillations}
A 50-MW data center is further connected at Bus 1 with the generator being set to supply 300 MW active power. Here, the data center operates at unity power factor and is subjected to a sinusoidal active power perturbation with a peak amplitude of 1 MW: $P_L(t)=\left[50+\sin\left(2\pi f_{\mathrm{osc}}t\right)\right]$ MW. The series-compensation level from $X_C$ is fixed at 30\%, and a 30-Mvar shunt capacitor provides local reactive power compensation at the load bus. The frequency response is evaluated using the original benchmark shaft damping data. EMT simulations are carried out in PSCAD/EMTDC by modulating the data center load at one forcing frequency per run and repeating the simulation over a set of selected frequencies.

Fig. \ref{fig:SBM_dc} (a) and (b) compare the predicted shaft torque responses with the EMT results. For the specified 1 MW data center oscillating at 24.7332 Hz, the largest shaft torque response (GEN--LP section) reaches up to 1.9481 p.u. from the predicted results and 1.6166 p.u. from the EMT results. These responses illustrate how a small data center load fluctuation can produce substantial shaft torque oscillations near resonance in the benchmark system.

\subsection{Application to the IEEE 39-Bus System}
\begin{figure}[!t]
\centering
\vspace{-0.1cm}
\includegraphics[width=3.2in]{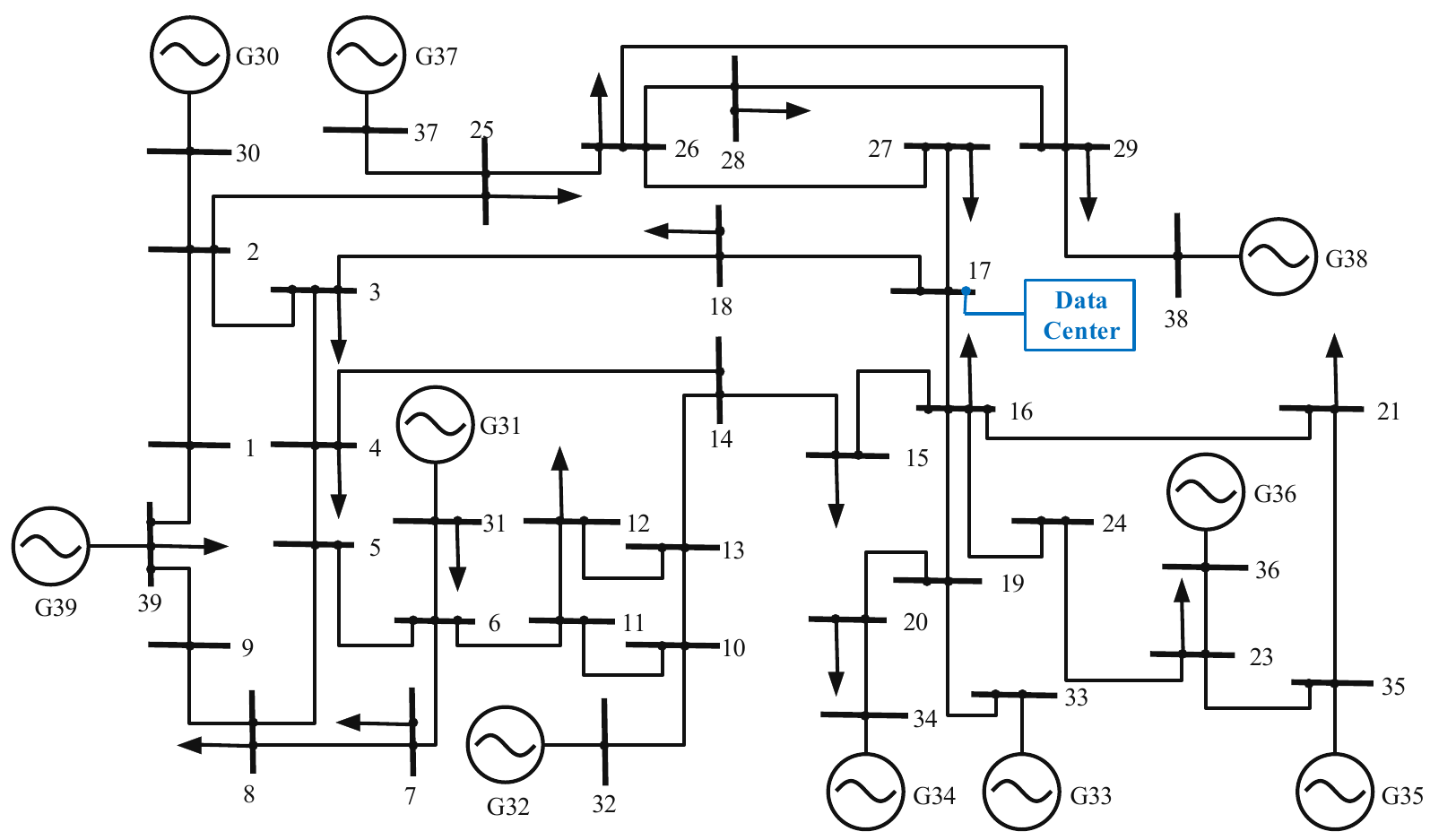}
\vspace{-0.3cm}
\caption{IEEE 39-bus test system with multi-mass generator models
and a data center connected at Bus 17.}
\vspace{-0.3cm}
\label{fig:IEEE39_sys}
\end{figure}

The proposed framework is further applied to the IEEE 39-bus test system shown in Fig. \ref{fig:IEEE39_sys}, with each of its ten generators represented by a multi-mass shaft model. The model parameters are drawn from data for three actual generating units provided by ISO-NE. Unit~1 data are assigned to G30, G33, G36, G39, Unit~2 data to G31, G34, and G37, and Unit~3 data to G32, G35, G38. All parameter data are scaled to the ratings of the corresponding generators. A data center is connected to Bus 17 imposing a sinusoidal active power perturbation, $\Delta P_{L,17}(t)=20\cos(2\pi f_{\mathrm{osc}}t)$ MW with no reactive power compensation.

\subsubsection{Generator terminal active power oscillation}
\begin{figure*}[t]
    \centering
    \includegraphics[width=\textwidth]{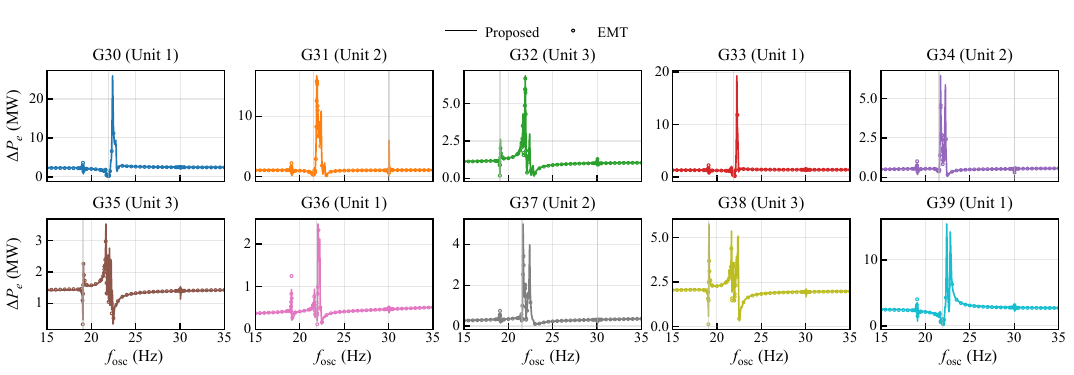}
    \caption{Generator terminal active power oscillation amplitude for $\Delta P_{L,17}(t)=20\cos(2\pi f_{\mathrm{osc}}t)$~MW imposed at Bus 17.}
    \label{fig:ieee39_Pe}
\end{figure*}

Fig. \ref{fig:ieee39_Pe} compares the amplitudes of the generator terminal active power oscillations induced by data center oscillation for different values of $f_\mathrm{osc}$ using the proposed framework and point-by-point EMT simulations. It shows that the proposed framework agrees closely with the EMT results over the entire 15--35 Hz range, including the sharp resonant peaks near the torsional modes. This range is chosen to cover all torsional modes of the test system. For other systems, it can be adjusted as needed to encompass their torsional modal frequencies.

The computations were performed on a workstation equipped with an AMD Ryzen 9 9950X processor (16 cores, 32 threads) and 64 GB of RAM, running 64-bit Windows 11 Pro. Following model preparation, the proposed method recorded a frequency-sweep time of 0.83 s for 565 frequency points over 15--35 Hz, whereas the sequential PSCAD sweep required 349.27 min for 101 frequency points over the same range. These results indicate a substantial computational advantage of the proposed method for frequency-response evaluation in this case study, despite its denser frequency grid.




\begin{figure*}[t]
    \centering
    \includegraphics[width=\textwidth]{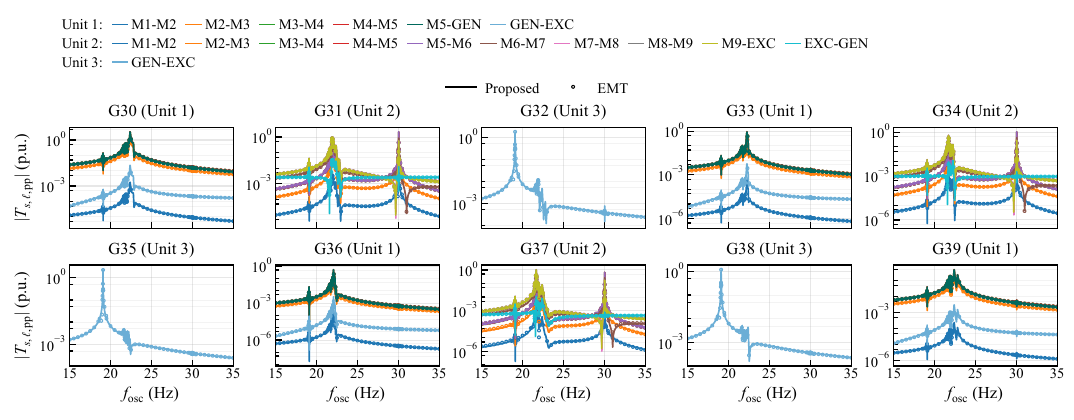}
    \caption{Generator shaft section torque peak-to-peak values for $\Delta P_{L,17}(t)=20\cos(2\pi f_{\mathrm{osc}}t)$~MW imposed at Bus 17.}
    \label{fig:ieee39_Ts}
\end{figure*}
\begin{table}[!t]
\centering
\caption{System Torsional Modes and Shaft Torque Responses}
\label{tab:ieee39_modal_shaft}
\footnotesize
\setlength{\tabcolsep}{3pt}
\renewcommand{\arraystretch}{1.05}
\begin{tabular*}{\columnwidth}{@{\extracolsep{\fill}}ccccccc@{}}
\toprule
Mode & $f_m$ (Hz) & $\sigma_m$ (s$^{-1}$) & $\zeta_m$ (\%)
& \shortstack{Dom.\\gen.} & Section
& \shortstack{$T_{s,\mathrm{pp}}^{\max}(f_m)$\\(p.u.)} \\
\midrule
T1  & 19.08307 & $-0.02118$ & 0.01767 & G35 & GEN--EXC & 2.3191 \\
T2  & 19.09039 & $-0.01361$ & 0.01134 & G38 & GEN--EXC & 0.9574 \\
T3  & 19.09251 & $-0.01174$ & 0.00979 & G32 & GEN--EXC & 0.8906 \\
T4  & 21.66309 & $-0.08708$ & 0.06398 & G37 & M7--M8   & 1.0503 \\
T5  & 21.70122 & $-0.10699$ & 0.07846 & G34 & M7--M8   & 0.6046 \\
T6  & 21.91746 & $-0.25321$ & 0.18387 & G31 & M7--M8   & 0.9296 \\
T7  & 22.06467 & $-0.07466$ & 0.05385 & G36 & M3--M4   & 0.5196 \\
T8  & 22.23892 & $-0.14185$ & 0.10152 & G33 & M3--M4   & 0.8972 \\
T9  & 22.38596 & $-0.23482$ & 0.16695 & G30 & M3--M4   & 3.4122 \\
T10 & 22.79377 & $-0.28490$ & 0.19893 & G39 & M3--M4   & 0.2191 \\
T11 & 30.04339 & $-0.00566$ & 0.00300 & G37 & M5--M6   & 0.6719 \\
T12 & 30.04424 & $-0.00676$ & 0.00358 & G34 & M5--M6   & 1.0553 \\
T13 & 30.04917 & $-0.00871$ & 0.00461 & G31 & M5--M6   & 2.1698 \\
\bottomrule
\end{tabular*}
\vspace{2pt}
\begin{minipage}{\columnwidth}\footnotesize
$\lambda_m=\sigma_m+{j}2\pi f_m$ and $\zeta_m=-\sigma_m/|\lambda_m|$.
\end{minipage}
\end{table}
\begin{figure}[!t]
\centering
\vspace{-0.1cm}
\includegraphics[width=3.3in]{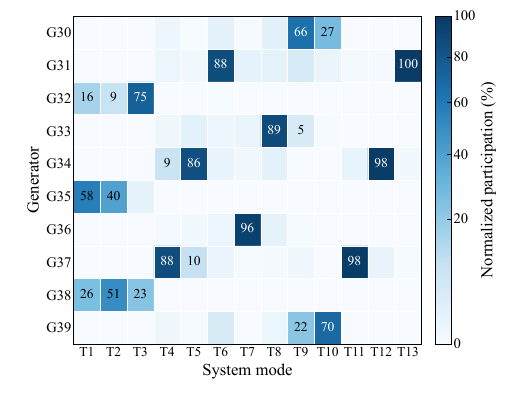}
\vspace{-0.3cm}
\caption{Normalized participation factors of generators G30–G39 in system torsional modes.}
\vspace{-0.3cm}
\label{fig:ieee39_pf}
\end{figure}
\begin{figure}[!t]
\centering
\vspace{-0.1cm}
\includegraphics[width=3.2in]{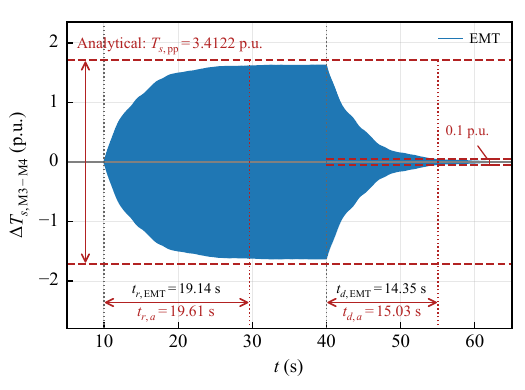}
\vspace{-0.3cm}
\caption{G30's M3-M4 shaft torque response to data center oscillation at $f_\mathrm{osc}=22.39$ Hz (data center connected at $t=10$ s and disconnected at $t=40$ s).}
\vspace{-0.3cm}
\label{fig:ieee39_G30}
\end{figure}
\subsubsection{Shaft torque response}
Fig. \ref{fig:ieee39_Ts} shows the peak-to-peak torque of every shaft section of the ten generators under the same forcing, $\Delta P_{L,17}(t)=20\cos(2\pi f_{\mathrm{osc}}t)$ MW at Bus 17. The proposed framework and the EMT simulations agree for all units, sections, and forcing frequencies over a response range spanning six orders of magnitude. The agreement also extends to the small features that each shaft exhibits at the torsional frequencies of the other unit types. The Unit 3 machines (G32, G35, G38) have a single torsional mode near 19.1 Hz, yet their GEN--EXC torque shows distinct features between 21.6 and 22.8 Hz, where the Unit 1 and Unit 2 modes lie, and the Unit 1 and Unit 2 machines in turn respond sharply at 19.1 Hz. This inter-machine interaction is transmitted through the network and is reproduced by the coupled formulation.

Table \ref{tab:ieee39_modal_shaft} lists the thirteen torsional modes of the system within 15--35~Hz and the peak-to-peak torque of the most stressed shaft section at each modal frequency $f_m$. Fig. \ref{fig:ieee39_pf} complements it with the participation of each generator in each mode, from which the dominant generator in Table \ref{tab:ieee39_modal_shaft} is identified. It can be seen that each natural frequency of a shaft model appears as a cluster of system modes, but these modes can exhibit markedly different distributions of generator participation. T1--T3 (Unit 3) are shared among G32, G35, and G38, indicating substantial joint participation of their shaft states. The Unit 2 modes, T4--T6 and T11--T13, are instead strongly localized in individual generators. Both behaviors occur within Unit 1, where T7--T8 are localized, and T9--T10 are shared mainly between G30 and G39. Accordingly, generators with the same shaft model need not exhibit the same participation pattern, and different modes associated with that model may involve either several machines or predominantly one machine. All thirteen modes are lightly damped, with $\zeta_m$ between 0.003\% and 0.2\%, and damping differs even among machines of the same type. The 30~Hz modes are the least damped, accounting for their very narrow peaks in Fig.~\ref{fig:ieee39_Ts}.

Specifically, Fig. \ref{fig:ieee39_G30} compares the EMT simulated M3--M4 shaft torque response of G30 with the proposed analytical estimates. The data center load, which injects a forced oscillation at $f_\mathrm{osc}=22.39$ Hz coinciding with torsional mode T9 in Table \ref{tab:ieee39_modal_shaft}, is connected at $t=10$ s and disconnected at $t=40$ s. During forcing, the maximum torque deviation obtained from the EMT simulation agrees closely with the analytical amplitude
$\Delta T_{s,\mathrm{M3-M4},\mathrm{pp}}=3.4122$ p.u. The build-up and decay of the torsional response are governed by the real part of the T9 eigenvalue, $\sigma_m=-0.23482$ s$^{-1}$. Under a single-mode approximation, the envelope of the forced response grows as $1-e^{\sigma_m (t-10)}$ and decays as $e^{\sigma_m (t-40)}$. Consequently, the time required to reach 99\% of the steady-state amplitude is approximately $t_{r,\mathrm{a}}=19.61$ s, which agrees with the EMT result $t_{r,\mathrm{EMT}}=19.14$ s. After data center disconnection, the evaluated time for $\Delta T_{s,\mathrm{M3-M4},\mathrm{pp}}$ to fall to 0.1 p.u. is about $t_{d,\mathrm{a}} = 15.03$ s, consistent with $t_{d,\mathrm{EMT}}=14.35$ s. Overall, the agreement confirms that the proposed method captures both the steady-state amplitude and the transient time scales of the forced torsional response.

Moreover, Table \ref{tab:ieee39_shaft_peak_power} shows the largest peak-to-peak shaft torque over 15--35~Hz, the section in which it occurs, the corresponding frequency $f_k^\star$, and the terminal power oscillation at that frequency. It can be seen that $f_k^\star$ lies within a few millihertz of a modal frequency in Table \ref{tab:ieee39_modal_shaft}, but not necessarily of the mode dominated by generator~$k$ itself. For G32, G38, and G39 the maximum torque occurs at a mode dominated by another machine of the same type (T1 for G32 and G38, T9 for G39) and is 1.2 to 2.1 times the torque at the machine's own dominant mode. It shows that evaluating the response only at each machine's dominant modal frequency would therefore underestimate its maximum torque. 


\begin{table}[!t]
\centering
\caption{Response at Maximum Shaft Torques}
\label{tab:ieee39_shaft_peak_power}
\setlength{\tabcolsep}{3pt}
\renewcommand{\arraystretch}{1.15}
\begin{tabular*}{\columnwidth}{@{\extracolsep{\fill}}ccccc@{}}
\toprule
Gen. & Shaft section
& $T_{s,\mathrm{pp}}^{\max}$ (p.u.)
& $f_k^\star$ (Hz)
& $|\Delta P_{e,k}|$ (MW) \\
\midrule
G30 & M3--M4 & 3.4196 & 22.38841 & 26.4861 \\
G31 & M5--M6 & 2.1699 & 30.04919 & 6.3825 \\
G32 & GEN--EXC & 1.7524 & 19.08290 & 6.4029 \\
G33 & M3--M4 & 0.8987 & 22.24024 & 20.3211 \\
G34 & M5--M6 & 1.0556 & 30.04427 & 1.4882 \\
G35 & GEN--EXC & 2.3292 & 19.08339 & 11.1090 \\
G36 & M3--M4 & 0.5207 & 22.06545 & 4.6556 \\
G37 & M7--M8 & 1.0532 & 21.66413 & 8.0857 \\
G38 & GEN--EXC & 1.1703 & 19.08255 & 8.6885 \\
G39 & M3--M4 & 0.4647 & 22.38694 & 15.9526 \\
\bottomrule
\end{tabular*}
\par\vspace{2pt}
\begin{minipage}{\columnwidth}
\footnotesize
$f_k^\star$ is the frequency within 15--35 Hz at which the largest shaft torque response among all shaft sections occurs.
\end{minipage}
\end{table}


\subsubsection{Data center active power oscillation limit}
\begin{figure}[!t]
\centering
\vspace{-0.1cm}
\includegraphics[width=3.1in]{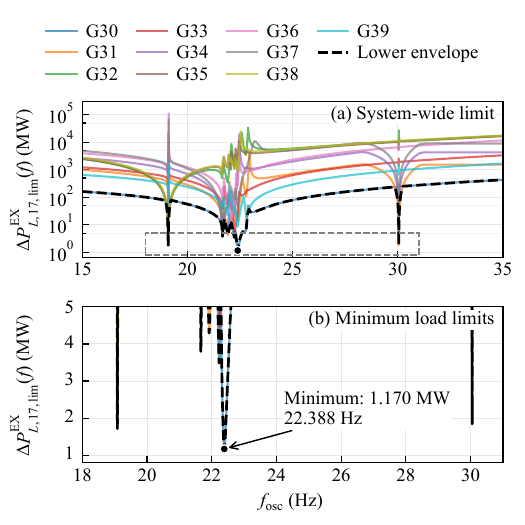}
\vspace{-0.3cm}
\caption{Admissible data center power oscillation amplitude at Bus 17 per generator and the system-wide limit for $EL_{\mathrm{set},k,\ell}=0.1$ p.u.}
\vspace{-0.3cm}
\label{fig:ieee39_plim}
\end{figure}

Fig.~\ref{fig:ieee39_plim} shows the derived admissible amplitude of the data center power oscillation at Bus 17 with respect to each generator, $\Delta P^{\mathrm{EX}}_{L,17,\mathrm{lim},k}(f)$ from \eqref{eq:limEX_section}, together with the system-wide limit $\Delta P^{\mathrm{EX}}_{L,17,\mathrm{lim}}(f)$ from \eqref{eq:limEX_system}, which is the lower envelope of the individual curves.

For $EL_{\mathrm{set},k,\ell}=0.1$ p.u., the system-wide admissible amplitude spans about two and a half orders of magnitude over 15--35~Hz. Away from the resonances it recovers to hundreds or thousands of megawatts, with G30 setting the lower envelope over most of the band. Near the torsional modes it falls below 10~MW, reaching its band-wide minimum of 1.17 MW at 22.39 Hz, where the M3--M4 section of G30 governs, with comparable minima near 19.08~Hz, set by the GEN--EXC section of G35, and near 30.05~Hz, set by the M5--M6 section of G31.

These results indicate that a frequency-independent limit may not be adequate across a wide range of oscillation frequencies. A limit set at the resonant minimum would be roughly two orders of magnitude more restrictive than necessary over most of the band, while a limit set at the off-resonance level would allow torques well beyond the endurance limit at the modes. 

\subsubsection{Impact of data center locations}
\begin{figure}[!t]
\centering
\vspace{-0.1cm}
\includegraphics[width=3.2in]{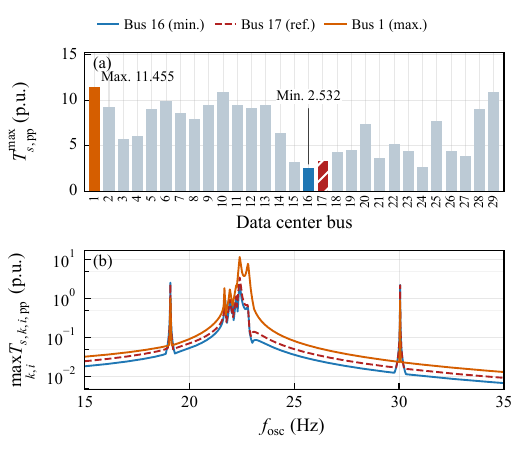}
\vspace{-0.3cm}
\caption{Impact of data center location on the maximum shaft torque. (a) Largest response per connection bus. (b) Upper envelopes of the shaft torques for the minimum, reference, and maximum buses.}
\vspace{-0.3cm}
\label{fig:ieee39_Lc}
\end{figure}

To isolate the impact of the connection location, the 20 MW data center perturbation is applied at each of Buses 1--29 in turn, and the frequency scan over 15--35 Hz is repeated for each bus. Fig.~\ref{fig:ieee39_Lc} (a) shows, for each candidate bus, the largest peak-to-peak shaft torque over all generators, shaft sections, and frequencies. Fig.~\ref{fig:ieee39_Lc} (b) shows the corresponding envelopes of the largest response over all generators and sections at each frequency for the buses yielding the minimum (Bus 16), the reference (Bus 17), and the maximum (Bus 1).

The results show that the system-wide maximum ranges from 2.532 p.u. at Bus 16 to 11.455 p.u. at Bus 1, with the Bus 17 reference being 3.420 p.u. The governing generator and section also change. Buses 17 and Bus 1 are governed by the M3--M4 section of G30 near 22.39 Hz, whereas Bus 16 is governed by the GEN--EXC section of G35 at 19.08 Hz. The connection location therefore affects both the magnitude and the distribution of torsional loading among generators and shaft sections, and the proposed framework allows the resulting torsional risk to be assessed rapidly for each candidate location.

\subsection{Application to the IEEE 39-Bus System with IBRs}
\begin{figure}[!t]
\centering
\vspace{-0.1cm}
\includegraphics[width=3.5in]{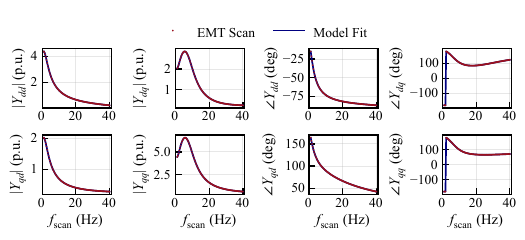}
\vspace{-0.3cm}
\caption{Frequency scan and model fit results of the black-box IBR at Bus 34.}
\vspace{-0.3cm}
\label{fig:ScanG30}
\end{figure}

The turbine-generator at Bus 34 is replaced by a grid-following inverter of the same rating \cite{Kenyon2021PSCADmodels}. The inverter is treated as a black-box IBR here and its small-signal $dq$ terminal admittance is identified through frequency scan via ModFit 26.0 provided by Powertech Labs at the operating point $P_0=510.76$ MW, $Q_0=234.04$ Mvar, and $V_0=1.08$~p.u. The frequency scan result for the IBR is shown in Fig. \ref{fig:ScanG30} and the fitted model is
\begin{equation*}
\bm{Y}_r(s)
=\bm{D}+\sum_{k=1}^{N_p}\frac{\bm{R}_k}{s-p_k},\qquad N_p=7,
\end{equation*}
where the poles comprise one real pole and three complex-conjugate pairs. Each pole is associated with a $2\times2$ residue matrix.

\begin{figure}[!t]
\centering
\vspace{-0.1cm}
\includegraphics[width=3.2in]{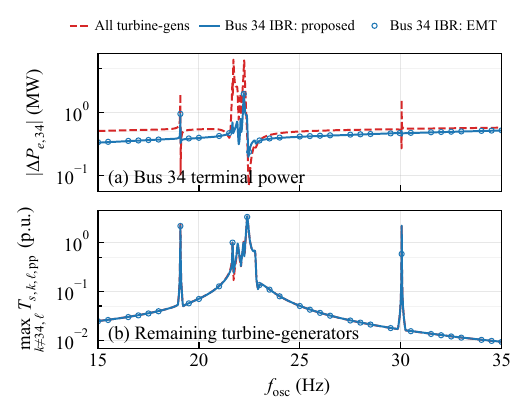}
\vspace{-0.3cm}
\caption{System response with G34 replaced by a black-box IBR.}
\vspace{-0.3cm}
\label{fig:ieee39_IBR}
\end{figure}

Table \ref{tab:ieee39_IBR} lists the torsional modes of the modified system. The two original T5 and T12 modes dominated by G34 in Table \ref{tab:ieee39_modal_shaft} disappear, and the remaining eleven modes are retained with frequency shifts of at most about 0.02 Hz. The changes in damping are larger but remain moderate. The mode of G37 at 21.67 Hz, in which G34 has a normalized participation of approximately 9\% in the original system, loses about 17\% of its damping, with $\sigma_m$ changing from $-0.087$ to $-0.073$~s$^{-1}$. The peak-to-peak torques at the modal frequencies change accordingly. Fig. \ref{fig:ieee39_IBR} (a) compares the power oscillation at Bus 34 with the turbine-generator and with the IBR in place. With the turbine-generator, the terminal power exhibits sharp resonant peaks and antiresonances at its own torsional modes and at those of the electrically close machines. With the IBR, these features are replaced by a smooth variation with $f_\mathrm{osc}$ on which only small features remain at the modes of the other machines. The proposed framework, which applied the fitted admittance of the IBR, agrees with the EMT simulation of the full IBR model over the entire band. Fig.~\ref{fig:ieee39_IBR} (b) shows the envelope of the peak-to-peak maximum shaft torques of the remaining nine turbine-generators. It is close to the original system, apart from a small change near 21.7 Hz where the G34 mode has been removed. 

The results confirm that the proposed framework could accommodate black-box IBRs through their terminal admittances and reproduce the coupled torsional response of the remaining machines without access to the internal information of the inverters.

\begin{table}[!t]
\centering
\caption{System Torsional Modes and Shaft Torque Responses With G34 Replaced by a Black-Box IBR}
\label{tab:ieee39_IBR}
\footnotesize
\setlength{\tabcolsep}{3pt}
\renewcommand{\arraystretch}{1.05}
\begin{tabular*}{\columnwidth}{@{\extracolsep{\fill}}ccccccc@{}}
\toprule
Mode & $f_m$ (Hz) & $\sigma_m$ (s$^{-1}$) & $\zeta_m$ (\%)
& \shortstack{Dom.\\gen.} & Section
& \shortstack{$T_{s,\mathrm{pp}}^{\max}(f_m)$\\(p.u.)} \\
\midrule
T1  & 19.08323 & $-0.02194$ & 0.01830 & G35 & GEN--EXC & 2.1882 \\
T2  & 19.09062 & $-0.01368$ & 0.01140 & G38 & GEN--EXC & 0.9476 \\
T3  & 19.09270 & $-0.01180$ & 0.00984 & G32 & GEN--EXC & 0.8811 \\
T4  & 21.66721 & $-0.07260$ & 0.05333 & G37 & M7--M8   & 1.0010 \\
T5  & 21.91600 & $-0.27142$ & 0.19711 & G31 & M7--M8   & 0.9429 \\
T6  & 22.06439 & $-0.08262$ & 0.05959 & G36 & M3--M4   & 0.5070 \\
T7  & 22.21867 & $-0.16139$ & 0.11561 & G33 & M3--M4   & 0.7985 \\
T8  & 22.38869 & $-0.23630$ & 0.16798 & G30 & M3--M4   & 3.3936 \\
T9  & 22.79726 & $-0.28560$ & 0.19939 & G39 & M3--M4   & 0.2178 \\
T10 & 30.04343 & $-0.00560$ & 0.00297 & G37 & M5--M6   & 0.6185 \\
T11 & 30.04921 & $-0.00879$ & 0.00465 & G31 & M5--M6   & 2.2028 \\
\bottomrule
\end{tabular*}
\end{table}

\section{Conclusion}
This paper proposes an analytical framework for assessing torsional oscillations of turbine-generators induced by periodic data center loads. A coupled generator–network model is established capturing the multi-mass shaft dynamics, the electromagnetic dynamics of the network, and the terminal admittances of black-box IBRs. By deriving explicit transfer functions from a data center power perturbation at a given bus to the generator states and shaft torques, the shaft response to arbitrary periodic load profiles is obtained through harmonic superposition. The derived frequency-dependent limits on data center power oscillations provide a quantitative approach to screen candidate connection buses against admissible shaft torque amplitudes. Validation on the IEEE SSR benchmark model and extensive case studies on the IEEE 39-bus system confirm that the proposed framework offers a reliable and computationally efficient alternative to EMT simulation for torsional screening of data center interconnections. Future work will extend the proposed framework to identify critical network impedance and to accommodate black-box multi-mass shaft models through hierarchical shaft modeling.

\section*{Acknowledgment}
The authors would like to thank Xi Lin and Andrew Musgrave of Powertech Labs Inc. for supporting this work by providing access to the DSAToolss\texttrademark{} Suite and implementing customized features.

\bibliographystyle{IEEEtran}
\bibliography{paper}

\end{document}